\documentclass{appolb}
\usepackage{graphicx}
\usepackage{wasysym}
\usepackage{pifont}
\usepackage{wrapfig}          
\usepackage{cite}
\usepackage{float}

\newcommand{\epo}{\;. }
\newcommand{\nn}{\nonumber}
\newcommand{\noi}{\noindent}
\newcommand{\amu}{a_\mu }
\newcommand{\amuh}{a_\mu^{\rm had} }
\newcommand{\mbo}[1]{$ #1 $ }
\newcommand{\power}[1]{\times 10^{#1} } 
\newcommand{\amm}{muon anomalous magnetic moment }
\newcommand{\bea}{\begin{eqnarray}}
\newcommand{\eea}{\end{eqnarray}}
\newcommand{\bit}{\begin{itemize}}
\newcommand{\eit}{\end{itemize}}

\newcommand{\cP}{{\cal P}}
\newcommand{\cQ}{{\cal Q}}

\newcommand{\cR}{{\cal R}}

\newcommand{\crn}{\nn \\}

\newcommand{\epm}{e^+e^-}
\newcommand{\amuexp}{a_\mu^{\mathrm{exp}}}
\newcommand{\amuthe}{a_\mu^{\mathrm{the}}}
\newcommand{\amuSM}{a_\mu^{\mathrm{SM}}}
\newcommand{\gv}{\mbox{GeV}}

\newcommand{\mv}{\mbox{MeV}}
\newcommand{\D}{\mathrm{d}}
\newcommand{\E}{\mathrm{e}}
\newcommand{\I}{\mathrm{i}}

\newcommand{\veps}{\varepsilon}
\newcommand{\semis}{\;;\;\;}

\newcommand{\bra}[1]{\langle{#1}|}

\newcommand{\ket}[1]{|{#1}\rangle}
\newcommand{\FF}{{\cal F}_{\pi^{0*}\gamma^*\gamma^*}}

\newcommand{\FFc}{{\cal F}_{\pi^{0*}\gamma^*\gamma}}
\newcommand{\FFbc}{{\cal F}_{\pi^{0*}\gamma\gamma}}
\newcommand{\FFac}{{\cal F}_{\pi^0\gamma^*\gamma}}

\newcommand{\FFabc}{{\cal F}_{\pi^0\gamma\gamma}}
\begin{document}
\title{%
\vskip-3cm{\baselineskip14pt
\centerline{\small DESY~18-057, HU-EP-18/11\hfill April 2018}}
\vskip1.5cm
The Muon g-2 in Progress%
\thanks{Presented at the XXIV Cracow EPIPHANY Conference on Advances in Heavy Flavour Physics,
9-12 January 2018, Crakow, Poland. Dedicated to the memory of Maria Krawczyk.%
To appear in Acta Physica Polonica B.
}%
}
\author{Fred Jegerlehner
\address{Humboldt-Universit\"at zu Berlin, Institut f\"ur Physik,
       Newtonstrasse 15,\\ D-12489 Berlin, Germany\\
Deutsches Elektronen-Synchrotron (DESY), Platanenallee 6,\\ D-15738 Zeuthen, Germany}
}
\maketitle
\begin{abstract}
Two next
generation muon $g-2$ experiments at Fermilab in the US and at J-PARC
in Japan have been designed to reach a four times better precision
from 0.54 ppm to 0.14 ppm and the challenge for the
theory side is to keep up in precision as far as possible. This has triggered
a lot of new research activities. The main motivation is the persisting 3 to 4 $\sigma$
deviation between standard theory and experiment. As Standard Model
predictions almost without exception match perfectly all other experimental
information, the deviation in one of the most precisely measured
quantities in particle physics remains a mystery and inspires the
imagination of model builders. Plenty of speculations are aiming to
explain what beyond the Standard Model effects could fill what seems
to be missing. Here very high precision experiments are competing with searches
for new physics at the high energy frontier lead by the Large Hadron
Collider at CERN. Actually, the tension is increasing steadily as no
new states are found which could accommodate the $g_\mu-2$
discrepancy. With the new muon $g-2$ experiments this discrepancy
would go up at least to 6 $\sigma$, in case the central values
do not move, up to 10 $\sigma$ could be reached if the present theory
error could be reduced by a factor of two. Interestingly, the new
$\alpha$ from Berkeley by R. H. Parker et al. Science 360, 191 (2018):
$\alpha^{-1}({\rm Cs18})=137.035999046(27)$ gives an $a_e$ prediction
$a_e=0.00115965218157(23)$ such that $a_e^\mathrm{exp}-a_e^\mathrm{the}=(-84\pm36)\power{-14}$
shows a $- 2.3~ \sigma$ deviation now. 
\end{abstract}
\section{Introduction}
A particle with spin \mbo{\vec{s}} like the muon exhibits a magnetic moment \mbo{\vec{\mu}}:
$$
\vec{\mu}={ g_\mu}\:\frac{e \hbar}{2m_\mu c}\:\vec{s}\;\; ;\;\;\;
{g_\mu=2\:(1+{ a_\mu})}\epo $$ Its Dirac value $g_\mu=2$ is modified by
radiative corrections $a_\mu=(g_\mu-2)/2=\frac{\alpha}{2\pi}+\cdots$ known as the
muon anomaly. The electromagnetic lepton vertex tested in the static
limit here is the simplest object you can think of.
\begin{figure}[h!]
\includegraphics[height=2.4cm]{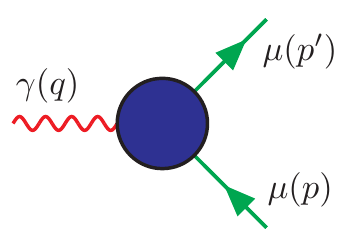}

\vspace*{-1.6cm}

\hspace*{4cm} $ =(-\I e)\:\bar{u}(p')\left[\gamma^\mu
F_1(q^2)+\I\,\frac{\sigma^{\mu\nu}q_\nu}{2m_\mu}F_2(q^2) \right]u(p)$
\end{figure}
\bea
F_1(0)=1\;\;;\;\;\; F_2(0)=a_\mu\epo
\eea
The muon anomaly $a_\mu$ is responsible for the Larmor spin precession
and for its tracking one needs polarized muons orbiting in a
homogeneous magnetic field. To this end one is shooting protons on a
target producing pions which decay by the parity violating weak process
$\pi^+ \to \mu^+ \nu_\mu$ into polarized muons of negative helicity
which are injected into a storage ring where they decay $\mu^+\to e^+
\nu_e \bar{\nu}_\mu$ producing positrons flying preferably in direction of the spin
of the decaying muon. For $\pi^-$ helicity and electron flight direction
are reversed. Indeed the two parity violating weak decays perfectly transport the
needed spin precession information.

The Larmor precession frequency \mbo{\vec{\omega}} developing in
the beam of polarized spinning muons injected into a homogeneous magnetic
field \mbo{\vec{B}} is detected by counting the positrons or
electrons ejected by the decaying muons preferably along the spin vector. 
\begin{figure}[h]
\includegraphics[width=0.47\textwidth]{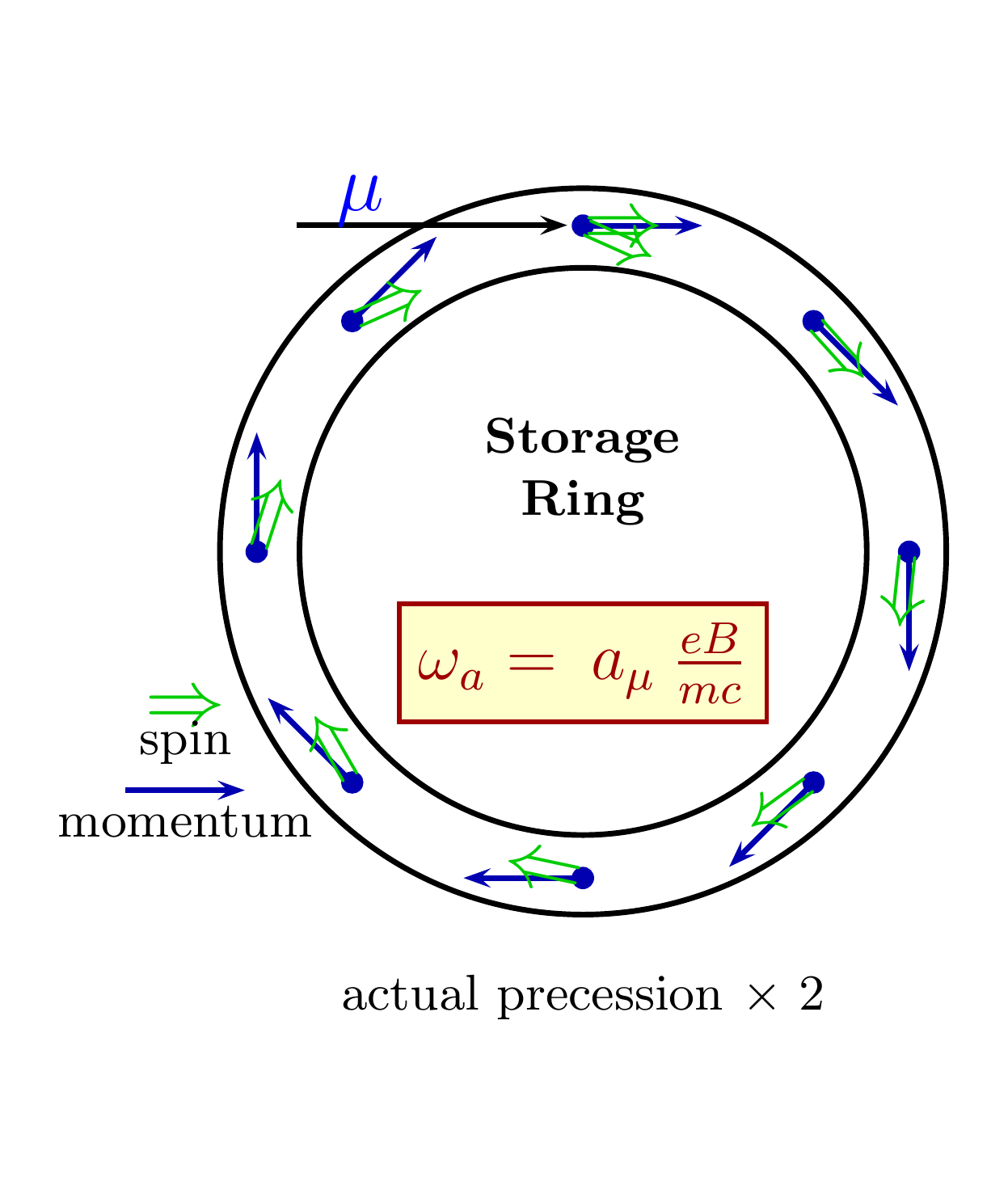}

\vspace*{-6cm}

\hspace*{0.47\textwidth}\includegraphics[width=0.53\textwidth]{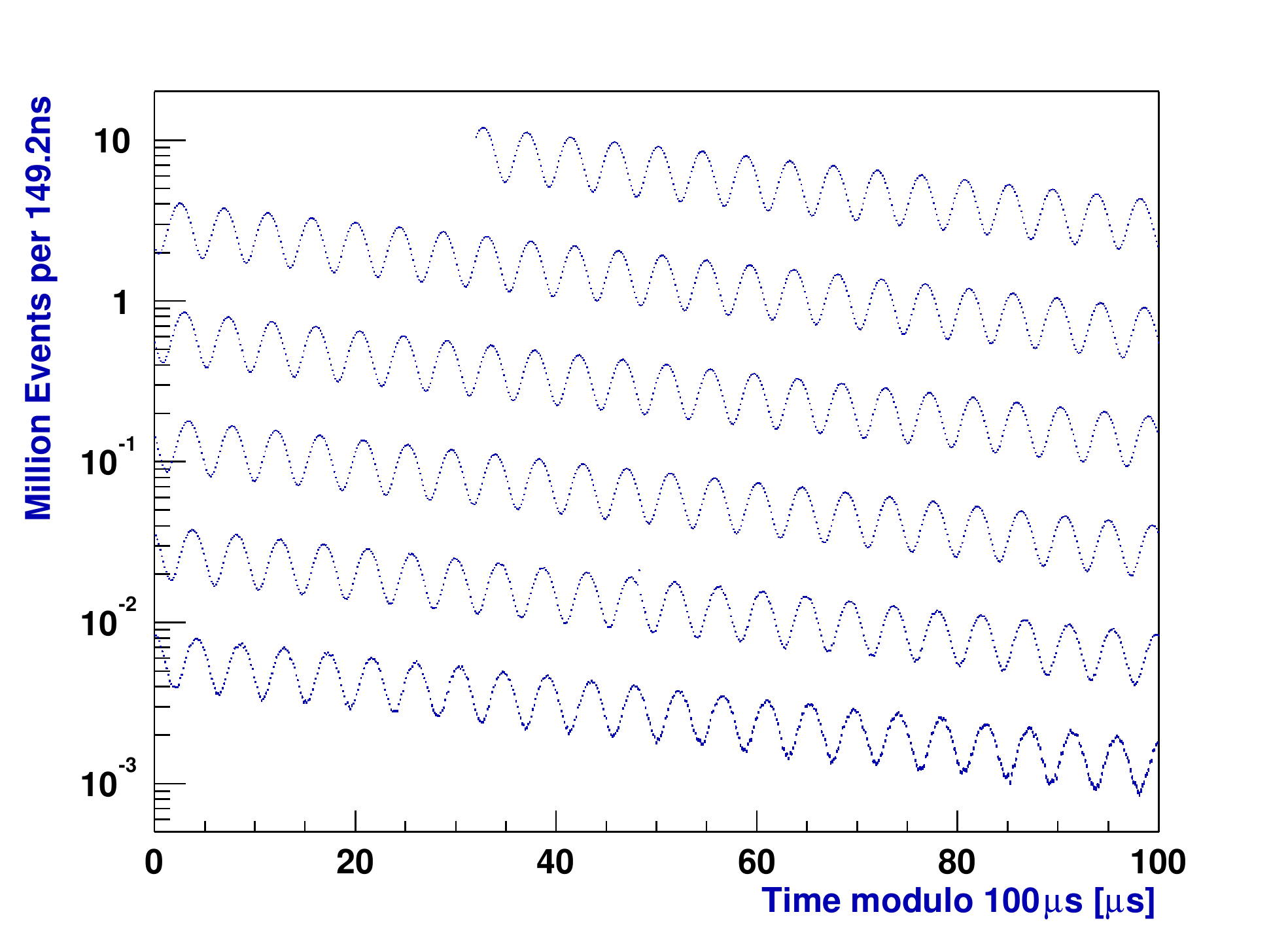}

\vspace*{2mm}

\caption{Left: the Larmor precession of a muon in a storage
ring. The spin is rotating $\sim 12'$ per circle. Right: the number of decay positrons with energy greater
than $E_{\rm th}$ emitted at time $t$ after muons are injected into
the storage ring is $ N(t)=N_0(E_{\rm th})\:\exp
\left(\frac{-t}{\gamma \tau_\mu}
\right)\:\left[1+A(E_{\rm th})\:\sin(\omega_a t+\phi(E_{\rm th})) \right]\,,
$
where $N_0(E_{\rm th})$ is a normalization factor, $\tau_\mu$ the muon life
time (in the muon rest frame), and $A(E_{\rm th})$ is the asymmetry factor for
positrons of energy $E>E_{\rm th}$. Courtesy of the E821 collaboration~\cite{Bennett:2006fi}.}
\label{fig:precessionwiggles}
\end{figure}
In storage ring type experiments as the CERN, Brookhaven and
Fermilab experiments the muon beam has to be focused by electric
quadrupole fields \mbo{\vec{E}}, but the beam dynamics can be kept
simple by running at the ``Magic Energy'' where \mbo{\vec{\omega}} is
directly proportional to
\mbo{\vec{B}}. At magic energy at about 3.1 GeV indeed we have
$$\vec{\omega}_a=\frac{e}{m}\left[ a_\mu
\vec{B}-\left(a_\mu-\frac{1}{\gamma^2-1} \right)\:
\vec{\beta} \times \vec{E}
 \right]_{{ \mathrm{at}\;"\mathrm{magic} \ \gamma"}}^{ E \sim 3.1 \gv}
\simeq \frac{e}{m}\left[a_\mu \vec{B} \right]\epo $$
First lepton magnetic moment measurements were by Stern and Gerlach in
1922 revealing the famous $g_e=2$ factor and much later by Kusch and Foley in 1948
who first observed the anomaly $g_e=2\:(1.00119\pm0.00005)\,$ for the
electron.

A crucial point is that at 3.1 GeV the muons life-time \mbo{\gamma
\tau_\mu} in the lab frame is by $\gamma \approx 29$ times longer than in the
rest frame. This makes it possible to store and let muons circle in a storage ring.

A precise experimental determination of $\amu$ has to be based 
on measurements of \textit{ratios of frequencies}.  
From \mbo{B=\frac{\hbar \omega_p}{2\mu_p}} and \mbo{\omega_a=\frac{e
a_\mu}{m_\mu c}\,B} and using
\mbo{\mu_\mu=\left(1+a_\mu\right)\,\frac{e\hbar}{2m_\mu c}} or
\mbo{\mu_\mu=\left(1+a_\mu\right)\,\frac{\hbar}{2}\,\frac{\omega_a}{a_\mu
B}=\left(\frac{1}{a_\mu}+1\right)\,\frac{\omega_a}{\omega_p}\,\mu_p}
and eliminating the muon mass one obtains \mbo{a_\mu=
\cR/(\lambda-\cR)} in terms of 3 frequency measurables:
\mbo{\tilde{\omega}_p=(e/m_\mu)\langle B
\rangle} the free proton NMR frequency\,, 
\mbo{\cR=\omega_a/\tilde{\omega}_p} the muon Larmor precession
frequency\,, and \mbo{\lambda=\omega_L/\tilde{\omega}_p=\mu_\mu/\mu_p}
from the muonium hyperfine
splitting experiment at LAMPF\,. 
The actual result from BNL ($\lambda$ updated) is~\cite{Bennett:2006fi}
$$\amuexp=(11\,659\,209.1\pm5.4\pm3.3[6.3])\power{-10}\epo$$

To come are two complementary experiments: the magic $\gamma$ improved
muon $g-2$ experiment at Fermilab, tuning
$\left(a_\mu-\frac{1}{\gamma^2-1} \right)=0$~\cite{Grange:2015fou},
and a novel cold muons experiment at J-PARC using a small storage ring
at $\vec{E}=0$~\cite{Mibe:2011zz}. Both experiments attempt to improve
the error by a factor
4. Most importantly, the ultra relativistic muons (CERN, BNL, Fermilab)
and the ultra cold muons (J-PARC) experiments exhibit very different systematics and the
latter will provide an important cross check of the magic gamma ones
(see~\cite{Hertzog:2015jru} and references therein). More on the
experimental aspects and status the reader may find in the contribution by Lusiani~\cite{Lusiani},
in these Proceedings.

The \amm is a number represented by an overlay of a large number of
individual quantum corrections of different sign, which depend on a
few fundamental parameters. In any renormalizable theory like the SM
it is an unambiguous prediction of that theory. It is an ideal monitor
for physics beyond the SM. The muon $g-2$ is about a
factor 19 or 46 (if theory uncertainties included) more sensitive to
New Physics (NP) than the electron $g-2\,,$  as we expect $\Delta a_\ell^{\rm
NP}=\alpha^{\rm NP}\,m_\ell^2/M_{\rm NP}^2$. The new muon $g-2$ search
for NP will take place as usual by confronting the new experiments
with SM theory\\[-4mm]
\bea
\Delta a_\mu^{\rm NP}  = \amuexp - \amuSM 
\eea
where\\[-8mm]
\bea
\amuexp= \frac{\omega_a/\tilde{\omega}_p}{\mu_{\mu}/\mu_p-\omega_a/\tilde{\omega}_p}
\eea
and\\[-6mm] 
\bea
\amuSM=a_\mu^{\rm QED}+a_\mu^{\rm weak} + a_\mu^{\rm
 HVP~LO}+a_\mu^{\rm HLbL}+a_\mu^{\rm HAD~HO}
\eea
and the goal is to reach a precision $\delta a_\mu^{\rm exp}\sim
140~{\rm ppb}$ in experiments and $\delta a_\mu^{\rm SM} < 220~{\rm
ppb}$ in theory. The coming round of ``digging deeper'' into the
virtual quantum world is based on an improvement of the 5 numbers that
have relevant uncertainties. These are $\omega_a$, $\tilde{\omega}_p$
and $\mu_{\mu}/\mu_p$, experimentally limited at 120 ppb.  The
expected experimental improvement will increase $ \Delta
\amu=\amuexp-\amuthe$ to $6.7\, \sigma$ if theory as today and to $ \Delta
\amu$ to $11.5\,\sigma$ if the SM prediction is improved by a reduction of
the hadronic uncertainty by a factor 2, which concerns  $a_\mu^{\rm
HVP~LO}$ and $a_\mu^{\rm HLbL}\,.$   That's what we hope to achieve! A
case that promises New Physics to be seen with high significance.

In the following I will focus on the parts of the SM prediction which are
limiting its precision, the leading order hadronic photon vacuum
polarization (LO-HVP) and the hadronic light-by-light (HLbL) contributions. 
\section{Evaluation of the Leading Order $a_\mu^{\rm had}$}
\begin{wrapfigure}{l}[3pt]{2.6cm}
\includegraphics[height=2cm]{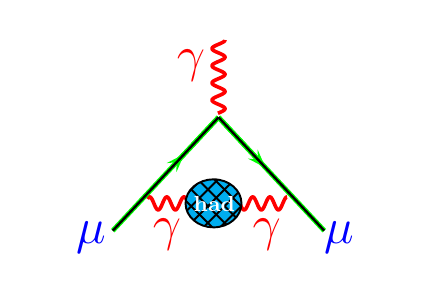}
\end{wrapfigure}
\begin{figure}[h]
\centering
\includegraphics[width=0.57\textwidth]{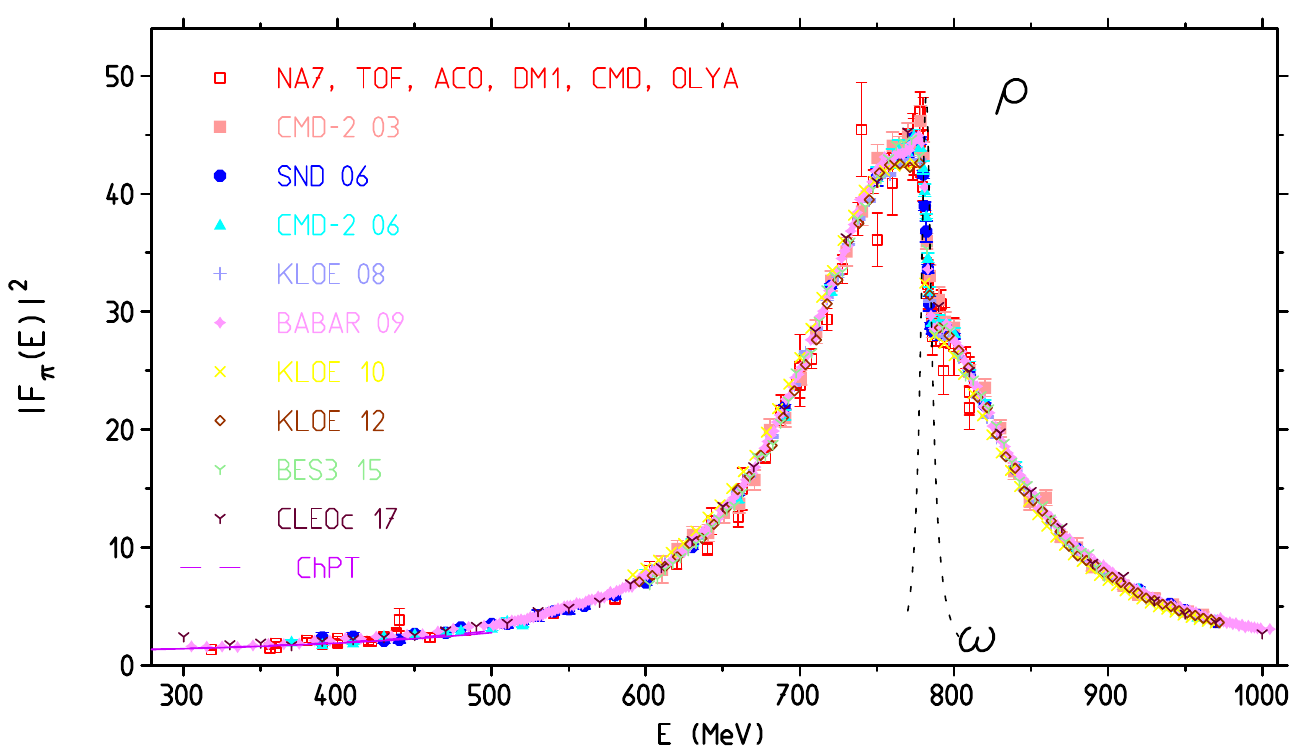}
\caption{A compilation of the modulus square of the pion form factor in the $\rho$ meson region,
which contributes about 75\% to $\amuh$. The corresponding $R(s)$ is
$R(s)=\frac14\,\beta_\pi^3\,|F_\pi^{(0)}(s)|^2\,,\,\,\beta_\pi=\sqrt{1=4m^2_\pi/s}$
is the pion velocity.}
\label{fig:VPdiadata}
\end{figure}
\begin{figure}
\centering
\includegraphics[height=2.9cm]{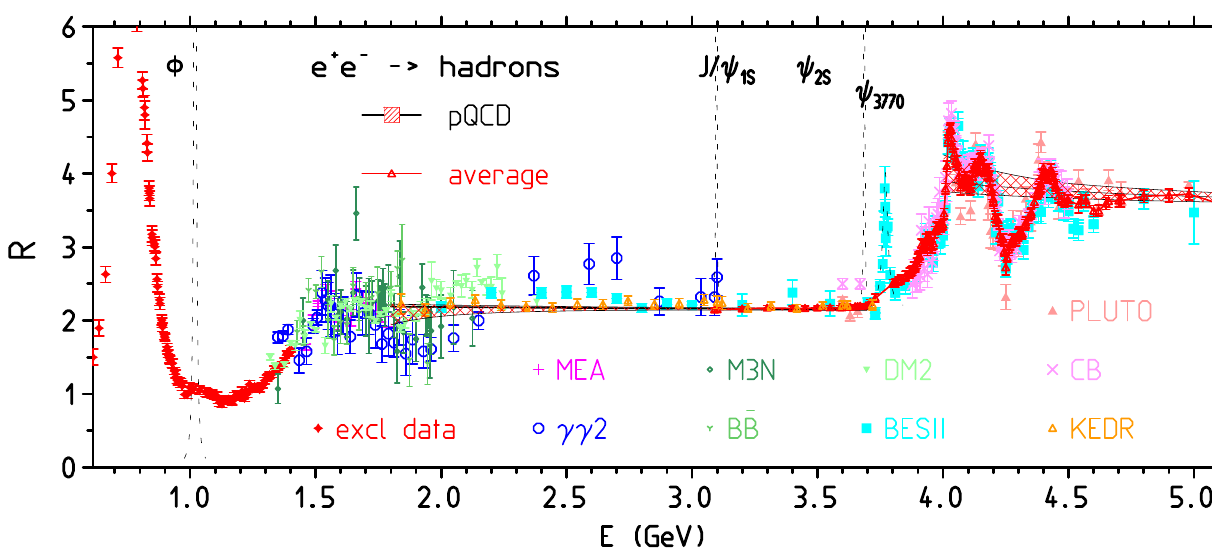}
\hspace*{-2mm}\includegraphics[height=2.9cm]{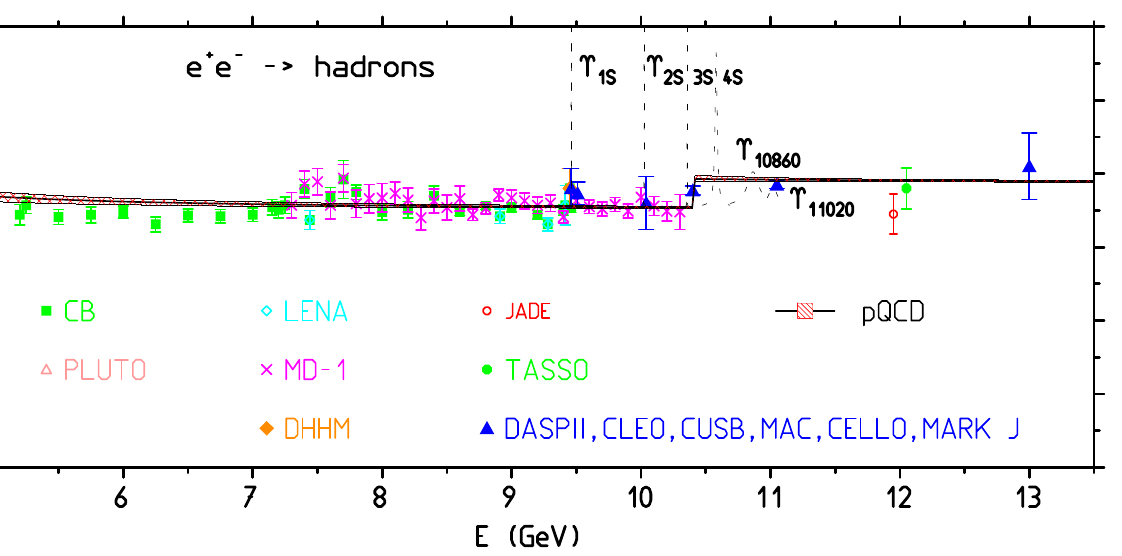}
\caption{The compilation of \mbo{R(s)}--data utilized.}
\label{fig:rofs} 
\end{figure}
\begin{figure}[!b]
\centering
\includegraphics[width=12.5cm]{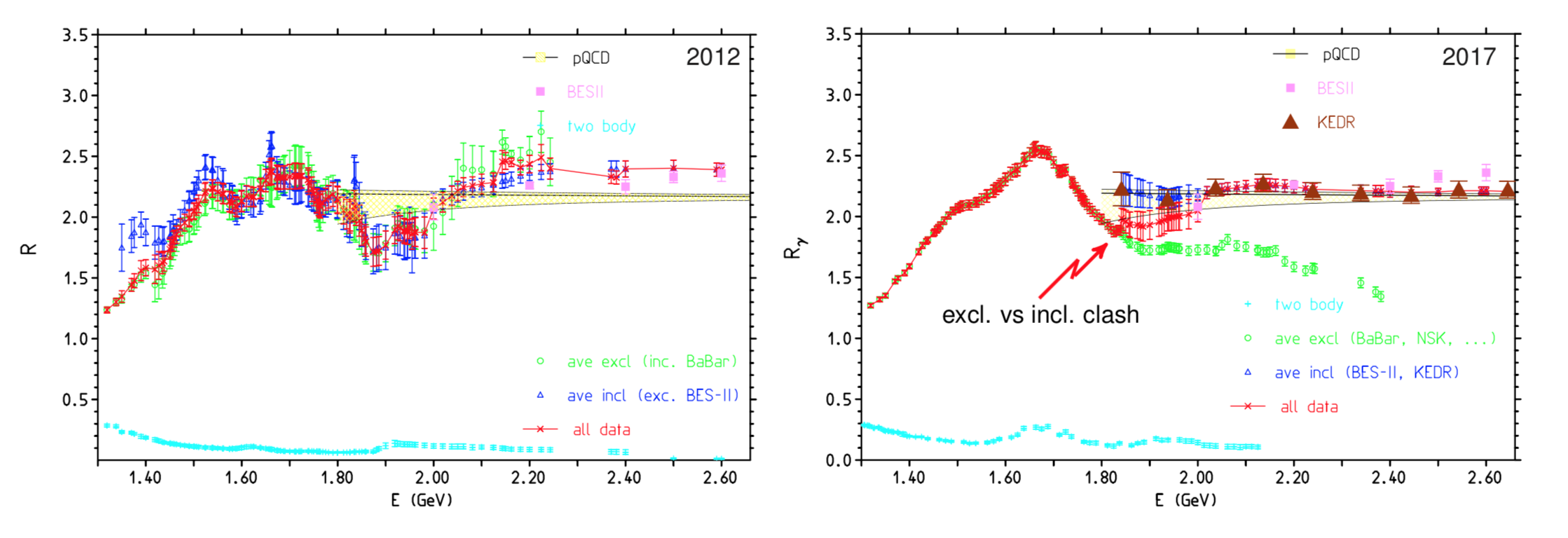}
\caption{Illustrating progress by BaBar and NSK exclusive channel data vs new
inclusive data by KEDR. Old Frascati (like $\gamma\gamma 2$) and Orsay
(DM2) data are superseded by much better BaBar data. The excl. data relative to
the pQCD band show an over shooting followed by an under shooting as
expected from quark-hadron duality. The KEDR point at 1.84 GeV seems
to violate duality expectations.}
\label{fig:exclvsincl}
\end{figure}

The hadronic contribution to the vacuum polarization can be evaluated,
with the help of dispersion relations (DR), from the energy scan of the
ratio $R(s)\equiv \sigma^{(0)}(e^+e^-\to \gamma^*\to {\rm
hadrons}) /\frac{4\pi\alpha^2}{3s}$ which can be measured up to some
energy $E_{\rm cut}$ above which we can safely use perturbative QCD
(pQCD) thanks to asymptotic freedom of QCD. Note that the DR requires the
undressed (bare) cross--section $\sigma^{(0)}(e^+e^-\to \gamma^*\to {\rm
hadrons})=\sigma(e^+e^-\to \gamma^*\to {\rm
hadrons})\,|\alpha(0)/\alpha(s)|^2$. The lowest order HVP
contribution is given by
\bea
\amuh = \left(\frac{\alpha m_\mu}{3\pi}
\right)^2 \bigg(\;\;\;
\int\limits_{m_{\pi^0}^2}^{E^2_{\rm cut}}ds\,
\frac{{ R^{\mathrm{data}}(s)}\;\hat{K}(s)}{s^2}
+ \int\limits_{E^2_{\rm cut}}^{\infty}ds\,
\frac{{R^{\mathrm{pQCD}}(s)}\;\hat{K}(s)}{s^2}\,\,
\bigg)\,,~~~~~~~~~~~~~
\label{amuDRbasic}
\eea
where $\hat{K}(s)$ is a known kernel function growing form $0.39,0.63\cdots$ at the
$m^2_{\pi^0},4m^2_\pi$ thresholds to 1 as $s\to \infty$. The integral is dominated by
the $\pi^+\pi^- \to \rho$ resonance peak shown in Fig.~\ref{fig:VPdiadata}. The
$R(s)$--data are displayed in Fig.~\ref{fig:rofs}. I apply pQCD from 5.2~GeV
to 9.46~GeV and above 11.5~GeV.
\begin{figure}
\centering
\includegraphics[height=4.5cm]{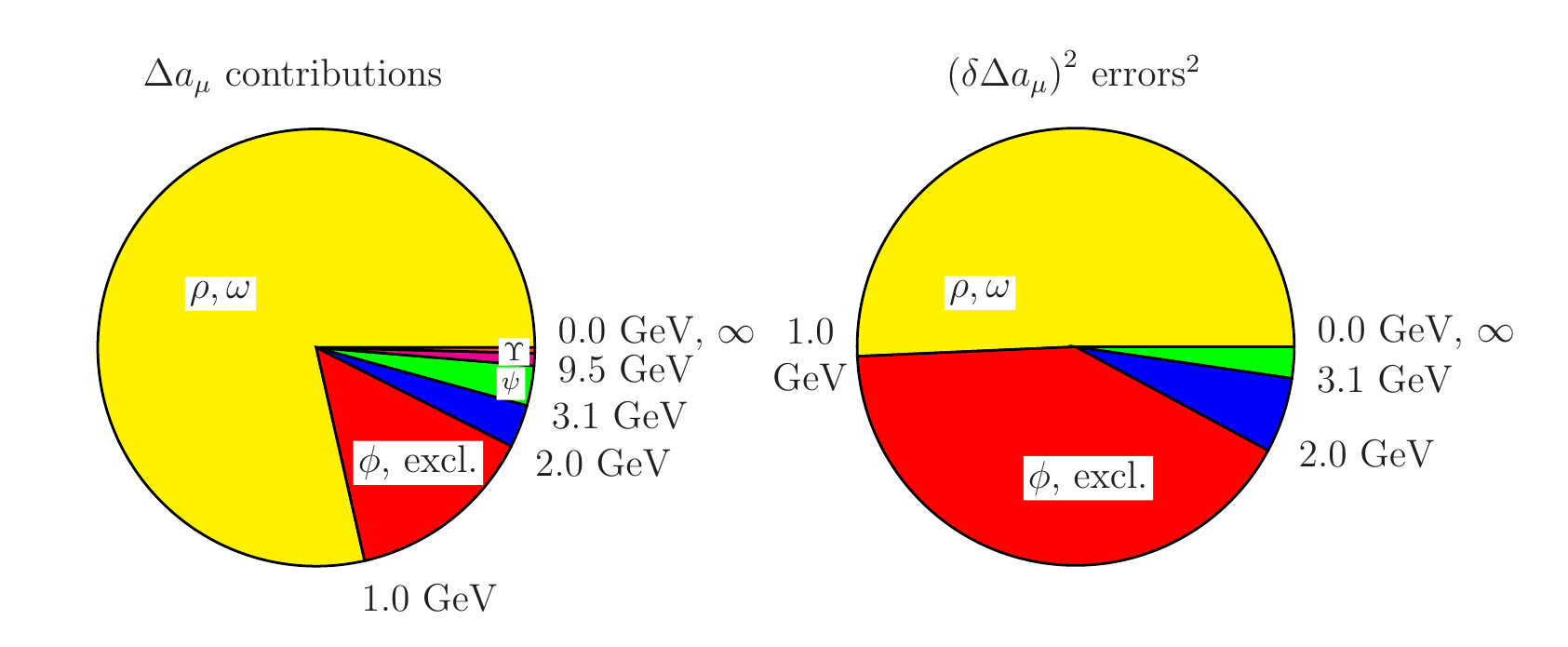}
\vspace*{-3mm}
\caption{Distribution of contributions and error squares from different
energy regions.}
\label{fig:gm2dist}
\end{figure}
The experimental errors imply the dominating theoretical
uncertainties. As a result I obtain~\cite{Jegerlehner:2017zsb,Jegerlehner:2017gek}
\begin{equation}
\amu^{\mathrm{had}}=(688.07\pm 4.14)[688.77\pm3.38]\:10^{-10}\semis
\epm-{\rm data \ based \ [incl.\ } \tau ]\,.
\label{LOHVP}
\end{equation}
Figure~\ref{fig:gm2dist} shows the distribution of contributions and
errors between different energy ranges.  One of the main issues is
$R(s)$ in the region from 1.2~GeV to 2.0~GeV (see
Fig.~\ref{fig:exclvsincl}), where more than 30 exclusive channels must
be measured and although it contributes about 14\% only of the total
it contributes about 42\% of the uncertainty.
In the low energy region, which is particularly important for the dispersive
evaluation of the hadronic contribution to the muon $g-2$, data have improved
dramatically in the past decade for the dominant $e^+e^- \to
\pi^+\pi^-$ channel (CMD-2~\cite{CMD203}, SND/Novosibirsk~\cite{SND06},
KLOE/Frascati~\cite{KLOE08,KLOE10,KLOE12,Anastasi:2017eio,Venanzoni:2017ggn},
BaBar/SLAC~\cite{BABARpipi}, BES-III/Beijing~\cite{BESIII}),
CLEOc/Cornell~\cite{Xiao:2017dqv} and the statistical errors are a
minor problem now. Similarly, the important region between 1.2 GeV to
2.4 GeV has been improved a lot by the BaBar exclusive channel
measurements in the ISR
mode~\cite{BaBar05,BaBar11,Davier:2015bka,Davier:2016udg}. Recent
data sets collected are: $e^+e^-\to 3(\pi^+\pi^-)$, $e^+e^-\to
\bar{p}p$ and $e^+ e^- \to K^0_{S}K^0_{L},K^+K^-$ from
CMD-3~\cite{Akhmetshin:2013xc,Kozyrev:2016raz},
and $e^+e^-\to \bar{n}n$, $e^+e^-\to \eta \pi^+\pi^-$, $e^+e^-\to
\pi^0\gamma$, $e^+e^- \to \omega\eta\pi^0$, $e^+e^- \to \omega\eta$,
$e^+e^- \to K^+K^-$ and $e^+e^- \to \omega\pi^0 \to \pi^0\pi^0\gamma$ from
SND~\cite{Achasov:2014ncd,Aulchenko:2014vkn,Achasov:2016bfr}.

\begin{figure}
\centering
\includegraphics[height=11cm]{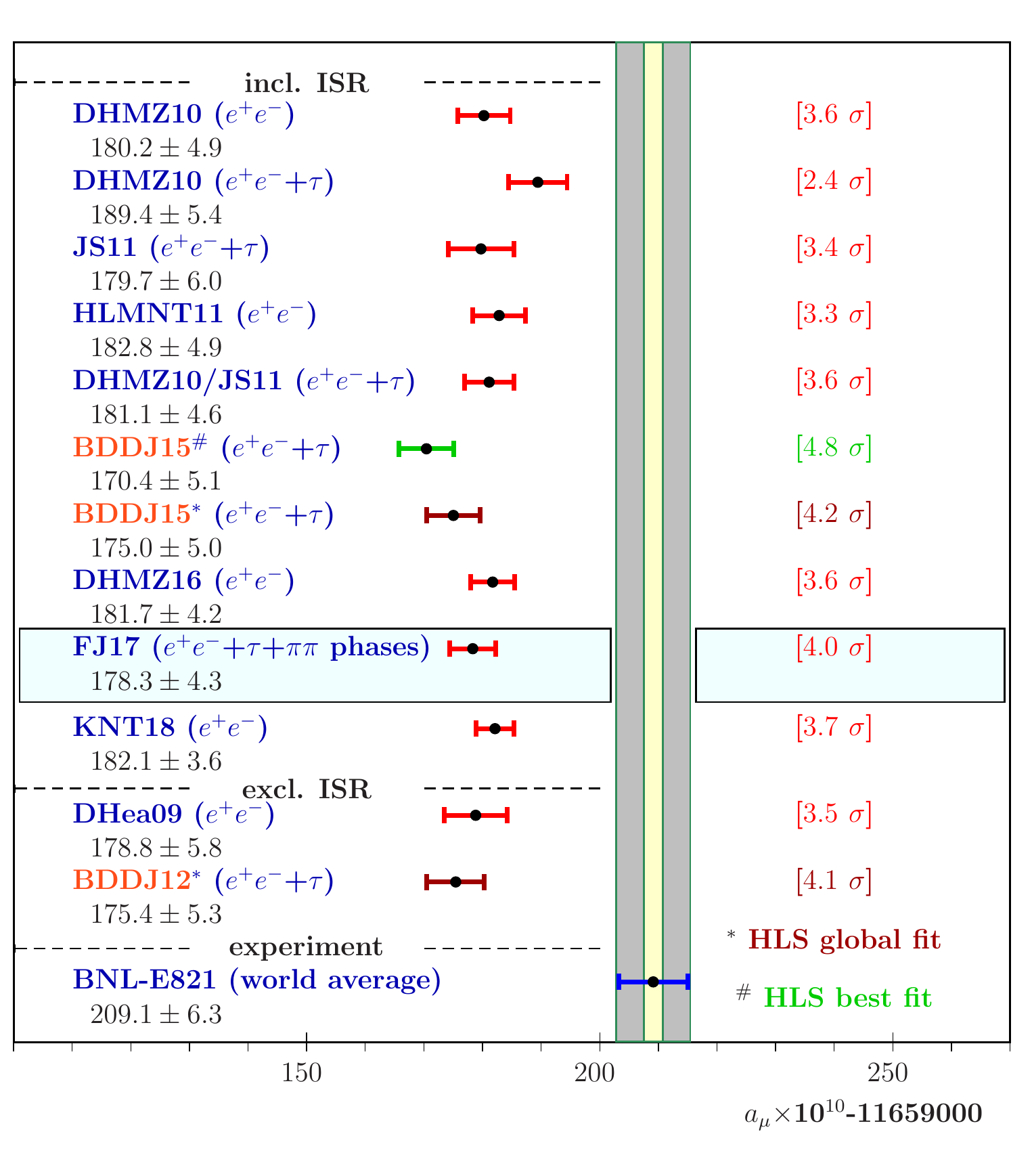}
\caption{Comparison with other results: DHMZ10~\cite{Davier:2010nc},
JS11~\cite{JS11}, HLMNT11~\cite{HMNT11},
BDDJ15~\cite{Benayoun:2015gxa}, DHMZ16~\cite{Davier:2016udg}, FJ17~\cite{Jegerlehner:2017zsb,Jegerlehner:2017gek}, 
DHea09~\cite{Davier:2009ag}, 
BDDJ12~\cite{Benayoun:2011mm}, KNT18~\cite{Keshavarzi:2018mgv}.
Two entries do not including IRS data. 
The narrow vertical
band illustrates the future precision expected.  
Note: results depend on which value is taken for HLbL. JS11 and BDDJ13
includes $116(39)\power{-11}$~\cite{JN} [JN]
others use $105(26)\power{-11}$~\cite{PdRV} [PdRV]. FJ17 includes
$\tau$ spectral data~\cite{JS11} and $\pi\pi$ scattering phase-shift data~\cite{Ananthanarayan:2013zua}.}
\label{fig:compare}
\end{figure}

Above 2~GeV fairly accurate BES-II data~\cite{BES02} are
available. A new inclusive determination of $R(s)$ in
the range 1.84 to 3.72 GeV has been obtained with the KEDR detector at
Novosibirsk~\cite{Anashin:2015woa} (see figures~\ref{fig:rofs} and~\ref{fig:exclvsincl}).
Recent new experimental input for HVP has been obtained by CMD-3 and
SND at VEPP-2000 via energy scan and by BESIII at PEPC in the ISR
setup. In Fig.~\ref{fig:compare} I show a collection of
results obtained by various groups since 2009. Figure~\ref{fig:compare} illustrates the
progress as well as the major uncertainties of SM predictions.  

\begin{figure}[t]
\centering
\includegraphics[width=0.48\textwidth]{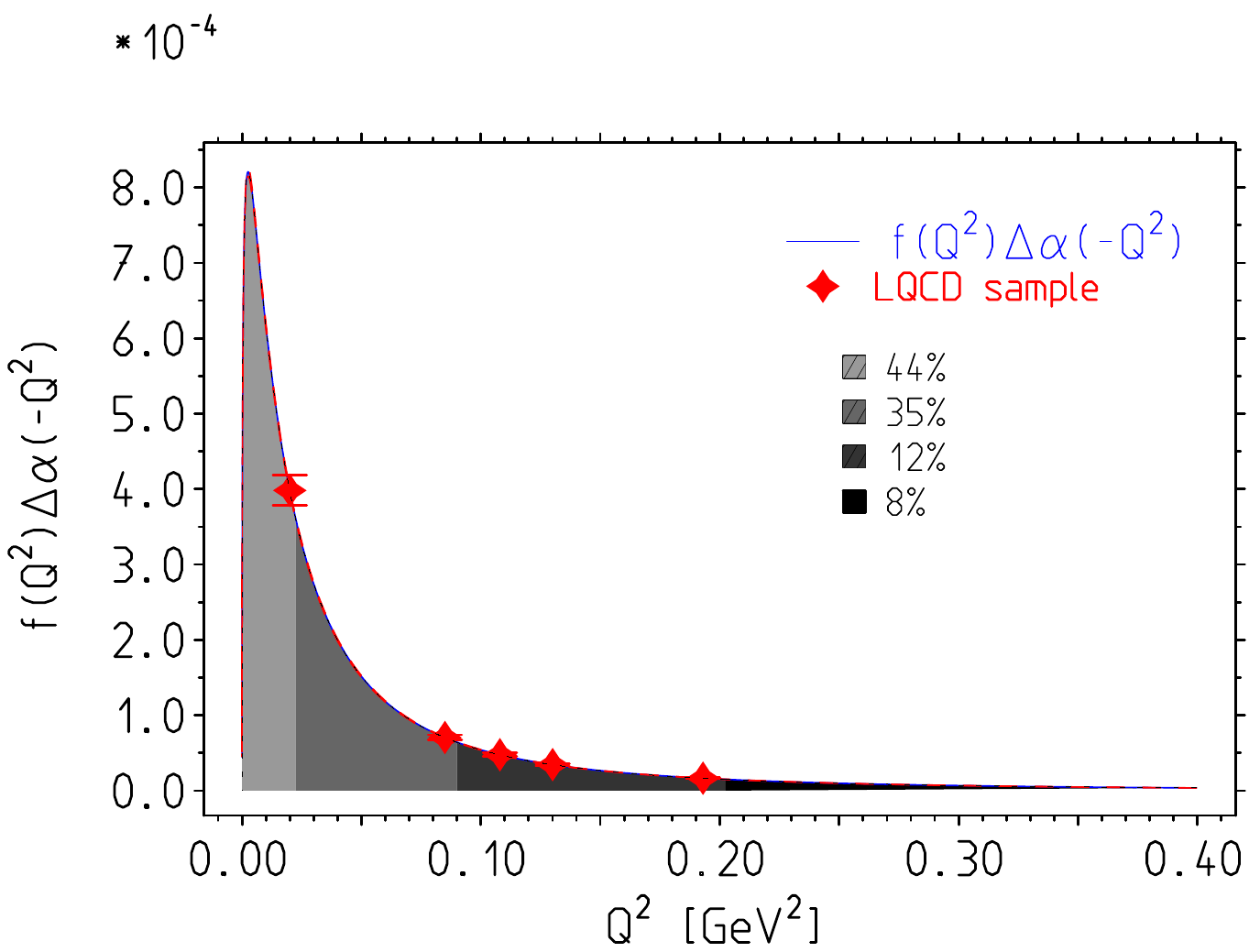}
\includegraphics[width=0.48\textwidth]{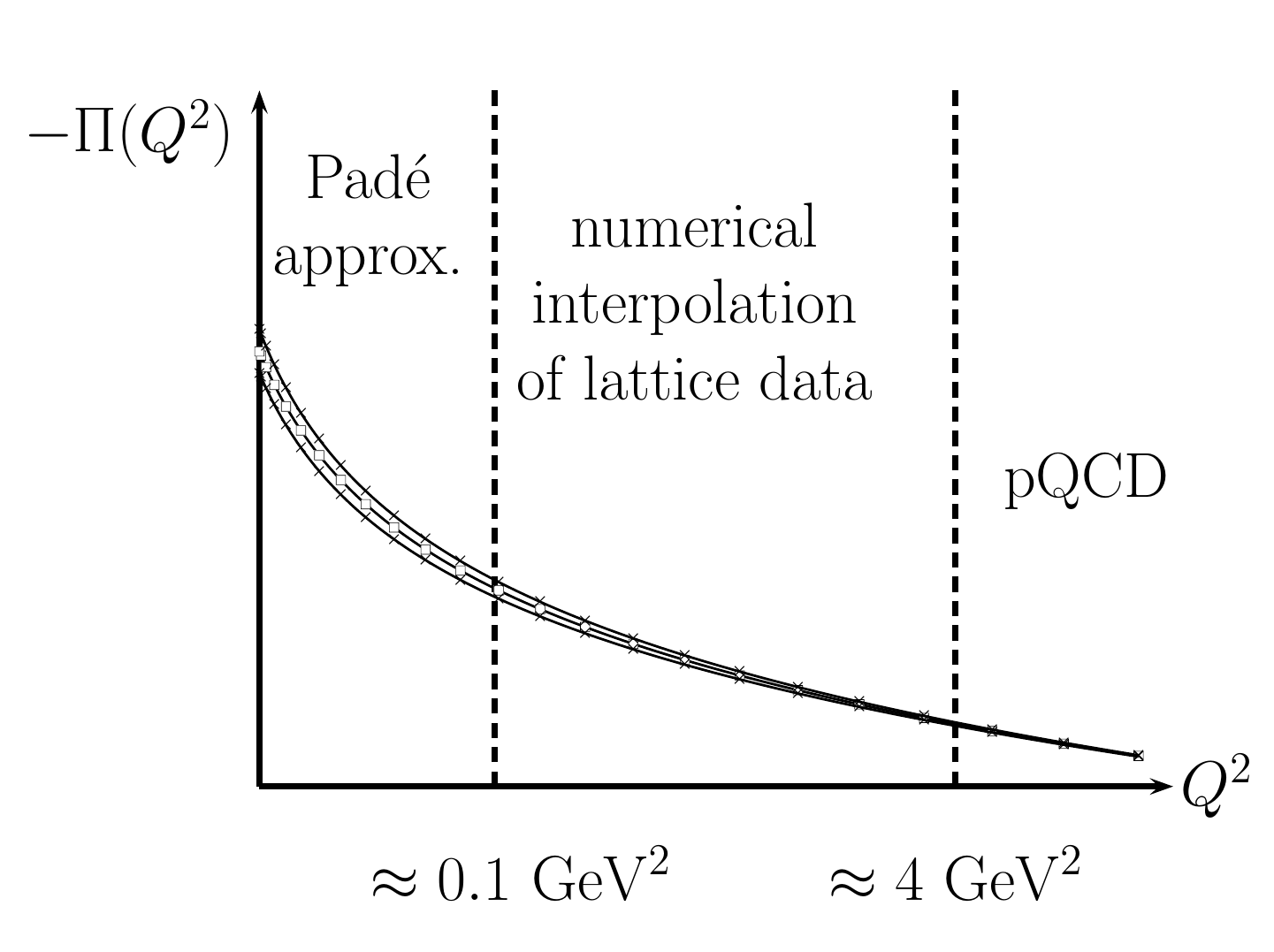}
\caption{Left: the $\amuh$ integrand as a function of $Q^2$. Ranges between $Q_i=0.00,\,0.15,\,0.30,\,0.45$ and $1.0~\gv$
and their percent contribution to $\amuh$. Right: range of direct lattice
data and the need for extrapolation.}
\label{fig:LQCDkern}
\end{figure}

\begin{figure}[h]
\includegraphics[width=0.65\textwidth]{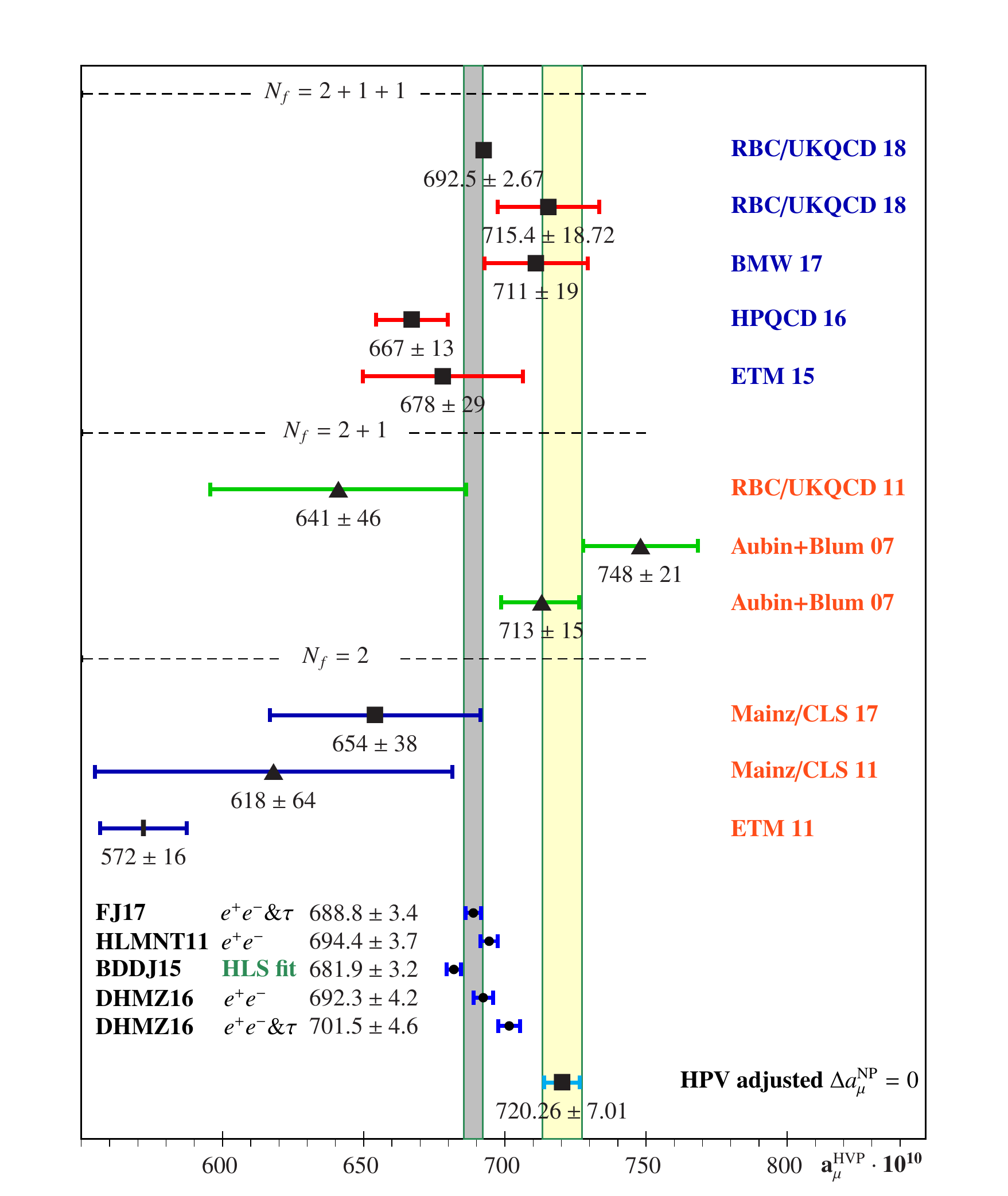}
    \begin{minipage}{0.33\textwidth}
\vspace*{-9.6cm}
\caption{\label{fig:LQCDx}
Summary of recent LQCD results~\cite{Blum:2018mom,Borsanyi:2017zdw,Chakraborty:2016mwy,Burger:2015hdi,Burger:2013jya,Boyle:2011hu,Aubin:2006xv,DellaMorte:2017dyu,DellaMorte:2011aa,Feng:2011zk}
for the leading order $a_\mu^{\rm HVP}$, in units $10^{-10}$. Labels:
\ding{110} marks $u,d,s,c$, \ding{115} $u,d,s$ and \ding{121} $u,d$
contributions. Individual flavor contributions from light $(u,d)$
amount to about 90\%, strange about 8\% and charm about 2\%.
The gray vertical band represents my evaluation. The wheat band
represents the HVP required such that theory matches the experimental
BNL result. Some recent $R$--data estimates are shown for comparison.}
    \end{minipage}
\vspace*{-7mm}
\end{figure}

Remarkable progress has been achieved by lattice QCD groups in
calculating $a_\mu^{\rm HVP}$. 
Primary object for HVP in LQCD is the electromagnetic current
correlator in Euclidean configuration space, which yields
the vacuum polarization function $\Pi(Q^2)$ needed to calculate
{\footnotesize $a_\mu^{\rm HVP}=4\alpha^2\int_0^\infty \D Q^2 f(Q^2)[\Pi(Q^2)-\Pi(0)]\,.$}
The integrand and the need for lattice size extrapolation is
illustrated in Fig.~\ref{fig:LQCDkern}.
Results are shown in Fig.~\ref{fig:LQCDx}.  The major part of LQCD
uncertainties comes from the need of extrapolations (finite volume,
lattice spacing and physical parametrers if not simulated at the
physical point). In fact the momentum region below $Q_{\rm
min}=2\pi/L$ ($L$ the lattice size) which for presently accessible
$Q_{\rm min}\sim 314~\mv$ accounts for about 40\% of
the $\amuh$ integral can only be obtained by extrapolation.  The very
precise RBC/UKQCD point is obtained by combining the directly
accessible lattice results only (33.5\%) with R--data (66.5\%) where
the latter are more precise.

\section{Hadronic Light-by-Light Contribution: Problems, Results}
Key object is the hadronic contribution to the full rank-four
light-by-light scattering tensor ($A^\mu(x)$ denoting the photon field)\\
\includegraphics[height=2.4cm]{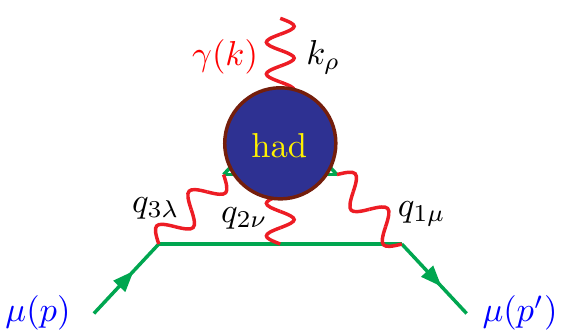}

\vspace*{-1.6cm}

\hspace*{4cm}{$\bra{0}T \{A^\mu(x_1) A^\nu(x_2) A^\rho(x_3)
A^\sigma(x_4)\} \ket{0}$}

\vspace*{1.2cm}

\noi
which embodies the four electromagnetic current amplitude
\bea
\Pi_{\mu\nu\lambda\rho}(q_1,q_2,q_3) &=&
\int \D^4x_1\: \D^4x_2\: \D^4x_3
\,\E^{\I\:(q_1x_1 + q_2x_2 + q_3 x_3)}\,
\nonumber\\
&&\quad
\times
\langle\,0\,\vert\,T
\{j_{\mu}(x_1)j_{\nu}(x_2)j_{\lambda}(x_3)j_{\rho}(0)\}  
\,\vert\,0\,\rangle \epo
\eea
The hadronic part with $j_\mu=j_\mu^{\rm had}=\frac23 \bar{u}\gamma_\mu u-\frac13
\bar{d}\gamma_\mu d - \frac13 \bar{s}\gamma_\mu s + \cdots$ shares the
following characteristic properties:
1) it is a non-perturbative object,
2) the covariant decomposition involves
138 Lorentz structures (43 gauge invariant),
3) 28 amplitudes can contribute to $g-2$, by permutation
      symmetry 19 thereof are independent,
4) fortunately HLbL is dominated by the pseudoscalar
exchanges $\pi^0,\eta,\eta'$
described by the effective Wess-Zumino Lagrangian,
5) generally, pQCD is used to
evaluate the short distance (S.D.) tail,
6) the dominant long
distance (L.D.) part must be evaluated using some low energy effective
model which includes the pseudoscalars as well as the vector mesons
(\mbo{\rho,\cdots}). The latter mediate the vector meson dominance
mechanism which is providing the necessary damping of the high energy
behavior. More recently, is has been shown that a data driven
dispersion relation approach is possible and very
promising~\cite{Colangelo:2015ama} and a number of improvements have
been obtained already~\cite{Pauk:2014rta,Colangelo:2017qdm}.

One usually applies appropriate low energy effective hadron theories,
like Hidden Local Symmetry (HLS), Extended Nambu-Jona-Lasinio (ENJL)
models, examples of the Resonance Lagrangian Approach (RLA), or large
\mbo{N_c} QCD inspired ans\"atze and other QCD inspired modelings which 
amount to calculate the following type of diagrams\\
\includegraphics[width=11.5cm]{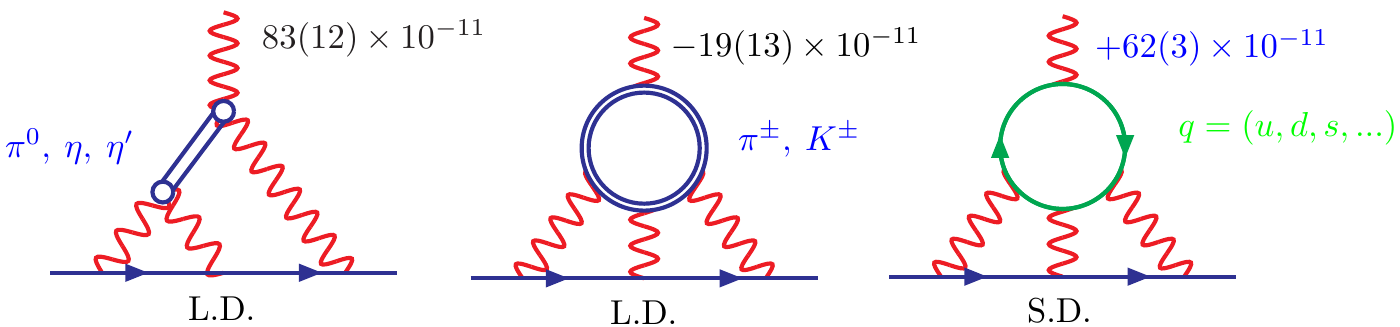}

\begin{figure}
\centering 
\includegraphics[width=0.495\textwidth]{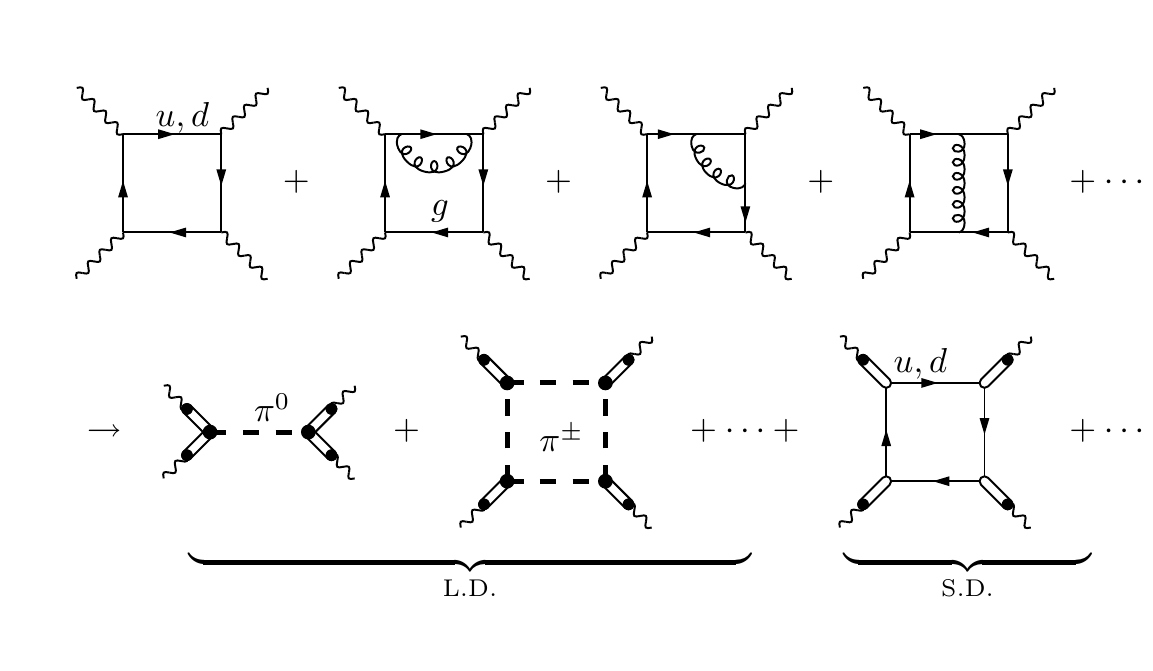}
\includegraphics[width=0.495\textwidth]{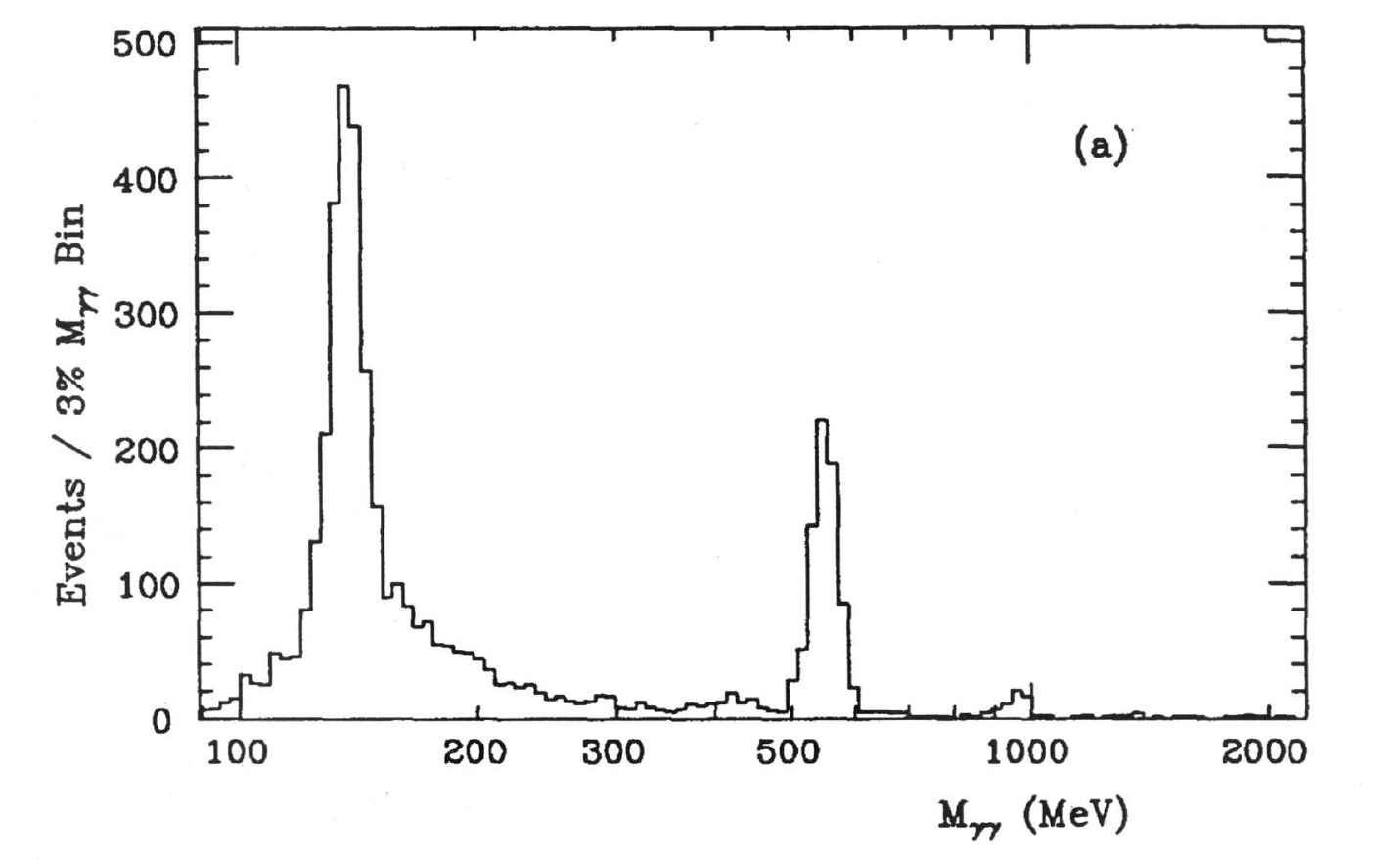}
\caption{Left: quark vs. hadron effective picture. Right: $\gamma
\gamma \to$ hadrons data [Crystal Ball 1988] show almost 
background free spikes of the pseudoscalar mesons!}
\label{fig:LbLfacts}
\end{figure}

\begin{figure}[!h]
\includegraphics[width=0.5\textwidth]{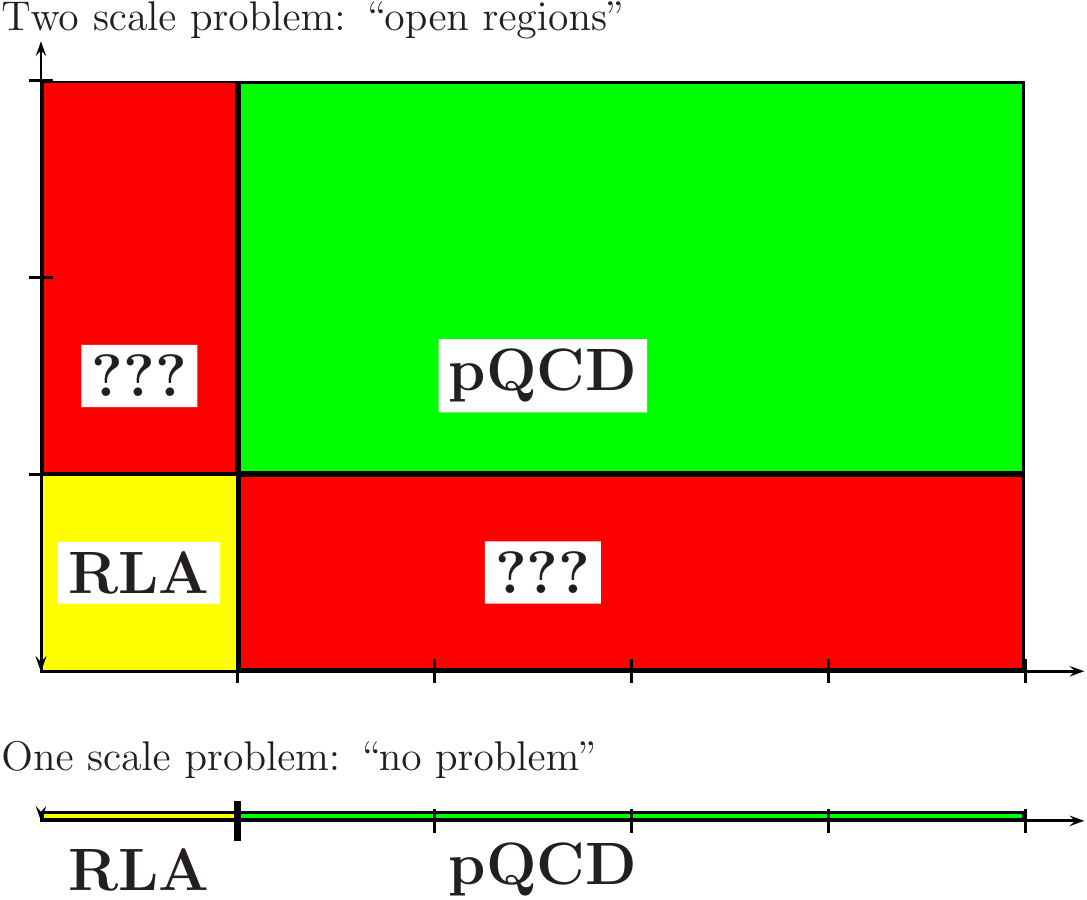}\small
\caption{The three scale HLbL exhibits not only L.D. and S.D. but also
mixed regions. Possible methods are listed to the right.}
\vspace*{-4.7cm}

\hspace*{6.4cm}\begin{tabular}{l}
??? multi-scale regions \\ 
  -- Data + Dispersion Relation,\\ 
  --  OPE, QCD factorization, \\
  -- Brodsky-Lepage approach  \\
  -- Models constrained by data\\ 
\end{tabular}\normalsize
\label{fig:multiscale}
\end{figure}

\noi
The non-perturbative L.D. contributions is dominated 
the \mbo{\pi^0} exchange and requires the knowledge of
the off-shell \mbo{\pi^0\gamma\gamma} form-factor (see Fig.~\ref{fig:LbLfacts}).
A basic problem in estimating the HLbL scattering contribution we have
because in contrast to the one-scale HVP, HLbL exhibits 3 different
energy scales. Fig.~\ref{fig:multiscale} illustrates the $(0,s_1,s_2)$--plane
of the general $(s,s_1,s_2)$--domain of the $\pi^0$ form-factor
$\FF(s,s_1,s_2)\,.$\\

\vspace*{2.7cm}

\begin{figure}[!h]
\centering
\includegraphics[width=0.35\textwidth]{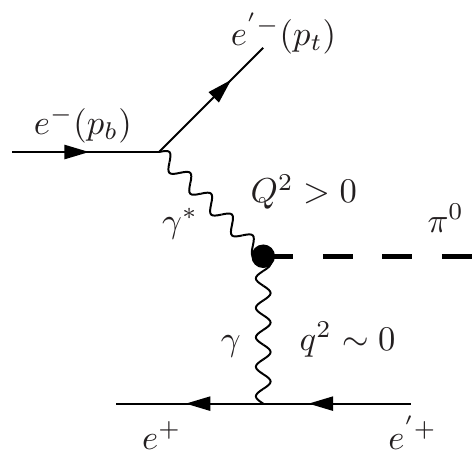}\hspace*{6mm}
\includegraphics[width=0.5\textwidth]{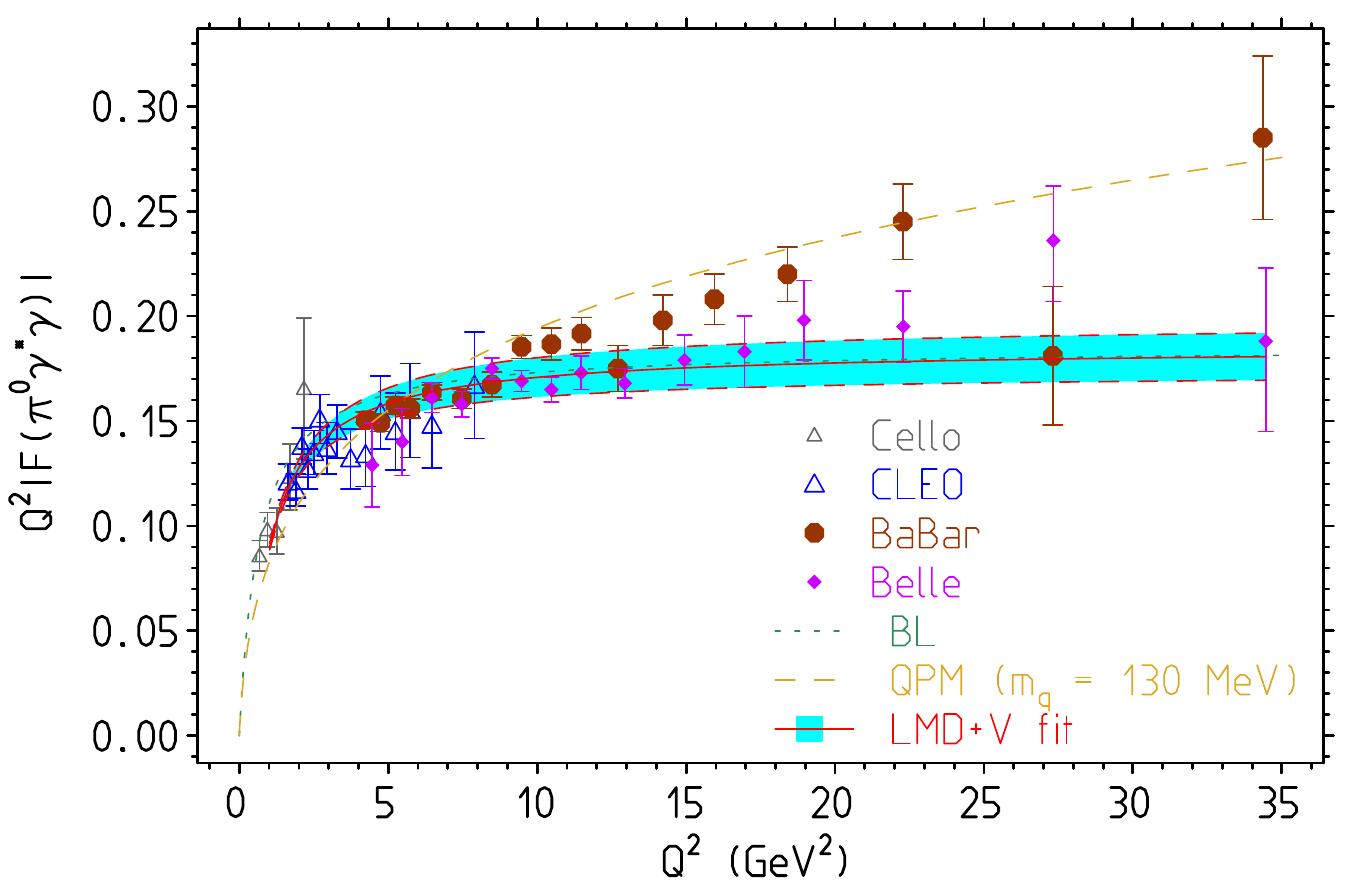}
\caption{CELLO, CLEO, BaBar and Belle measurements of the $\pi^0$ form factor $\FFac(m_\pi^2,-Q^2,0)$ at high 
space--like $Q^2$. Towards higher energies BaBar is somewhat
conflicting with Belle. The latter conforms with theory expectations,
which we use as an OPE constraint. More data are available for $\eta$
and $\eta'$ production.}
\label{fig:pi0ffexp}
\end{figure}

Lets focus on the leading $\pi^0$ exchange contribution. What do we know?\\
Constraint I: $\pi^0\to\gamma\gamma$ decay
\bit
\item
The constant $e^2\,\FFabc(m_\pi^2,0,0)=\frac{e^2 N_c}{12 \pi^2 f_\pi}
=\frac{\alpha}{\pi f_\pi} \approx 0.025\gv^{-1}$
is well determined by the { $\pi^0 \to \gamma \gamma$} decay rate
(from Wess-Zumino (WZ) Lagrangian). 
\item Information on $\FFac(m_\pi^2,-Q^2,0)$ come
from $\epm \to \epm \pi^0$ experiments as shown in Fig.~\ref{fig:pi0ffexp}. 
\eit

\noi
Constraint II: by the VMD mechanism the related Brodsky-Lepage
behavior {\mbo{\FFac(m_\pi^2,-Q^2,0) \simeq \frac{1}{4\pi^2
f_\pi}\frac{1}{1+(Q^2/8\pi^2 f_\pi^2)} \sim \frac{2f_\pi}{Q^2}}}
provides the necessary damping (cutoff) in order to obtain finite
integrals (the constant WZ form factor leads to a divergent
result). Variants of models satisfying the constraints I and II yield
similar answers. But ambiguities remain as only single tag data are
available (one photon real) so far, as displayed in
Fig.~\ref{fig:pi0ffexp}.

Recently the leading pseudoscalar meson exchange matrix element 
\bea
M_{\mu\nu}=\veps_{\mu \nu \alpha \beta} q_1^\alpha q_2^\beta
\FFbc(m_\pi^2; q_1^2,q_2^2)\,,
\eea
has been evaluated beyond the single tag case in lattice
QCD~\cite{Gerardin:2016cqj,Asmussen:2018ovy}. For the first time \mbo{\FFbc(m_\pi^2;
-Q^2,-Q^2)} could be measured on the lattice and clearly discriminates
all simple VMD model ans\"atze! What remains is the large--\mbo{N_c}
QCD (OPE constrained) LMD+V ansatz~\cite{KnechtNyffeler01}
\bea
\FF^{\rm LMD+V}(p_\pi^2, q_1^2, q_2^2)&=&
 \frac{F_\pi}{3}\,\frac{\cP(q_1^2,q_2^2,p_\pi^2)}{\cQ(q_1^2,q_2^2)} , \crn
\cP(q_1^2,q_2^2,p_\pi^2)&=&
{ h_0}\,q_1^2\,q_2^2\,(q_1^2 + q_2^2 + p_\pi^2) + { h_1}\,(q_1^2+q_2^2)^2
+ { h_2}\,q_1^2\,q_2^2\crn && + { h_3}\,(q_1^2+q_2^2)\,p_\pi^2 + { h_4}\,p_\pi^4
+{ h_5}\,(q_1^2+q_2^2) + { h_6}\,p_\pi^2 + { h_7}
, \crn
\cQ(q_1^2,q_2^2) &=&
({ M_{V_1}^2}-q_1^2)\,({ M_{V_2}^2}-q_1^2)\,({ M_{V_1}^2}-q_2^2)\,({ M_{V_2}^2}-q_2^2) ,
\eea
which for the pion-pole approximation $ p_\pi^2 = m_\pi^2$ is well constrained now, 
i.e. parameters \mbo{h_i\:(i=0,\cdots,7)} are rather well under
control by QCD asymptotics and experimental and lattice data. 
QCD + constraints by data fixes \mbo{h_0=-1}, \mbo{h_1=0}, 
\mbo{h_3,\,h_4,\,h_6} are absent in chiral limit such that
only \mbo{h_2,\,h_5} and \mbo{h_7} remain as essential parameters if
one adopts the VMD mechanism and identifies \mbo{{ M_{V_1}},\,{ M_{V_2}}}  with
\mbo{\rho,\,\rho'} masses.

\begin{figure}
\centering
\includegraphics[height=4.5cm]{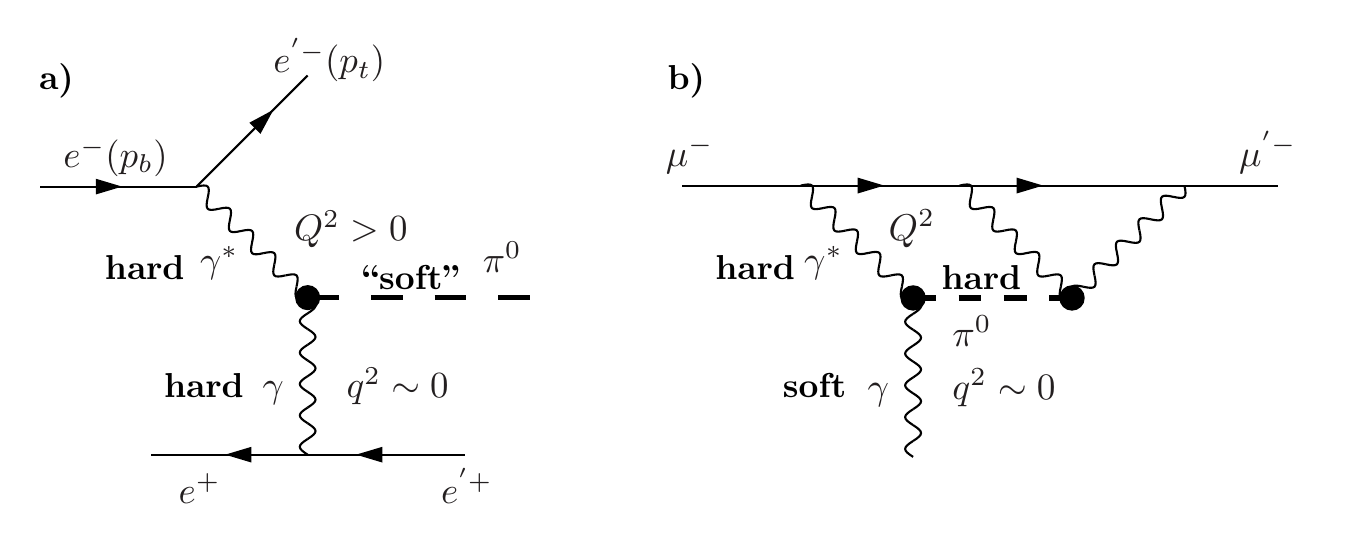}
\caption{Measured is $\FFac(m_\pi^2,-Q^2,0)$ at high 
space--like $Q^2$, needed at external vertex of the $g-2$ diagram is $\FFc(-Q^2,-Q^2,0)$
or $\FFc(q^2,q^2,0)$ if integral to be evaluated in Minkowski space.}
\label{fig:pi0ffcontinued}
\end{figure}

One important issue concerns the need of analytic continuation, as
illustrated in Fig.~\ref{fig:pi0ffcontinued}.
In principle this should be answered within the dispersive approach to
HLbL or in lattice QCD (see
e.g.~\cite{Blum:2015gfa,Green:2015sra,Asmussen:2018ovy}).  So far most estimates
adopt the pion-pole approximation (except~\cite{Nyffeler:2009tw,JN})
and apply VMD dressing at external vertex (except~\cite{MV03}). Adopting a LMD+V fit, my
estimation for the leading LbL contribution from PS mesons is\\
\centerline{\mbo{\amu[\pi^0,\eta,\eta'] \sim ({95.45{ =[64.68+14.87+15.90]}\pm 12.40}) \power{-11}\epo}}
Table~\ref{tab:pi0res} lists a number of results for the \mbo{\pi^0}-exchange
contribution using very different approaches.
\begin{table}
\centering
\caption{Some results for the leading \mbo{\pi^0}-exchange
contribution to the HLbL.}
\label{tab:pi0res}
\begin{tabular}{lcl}
\hline
\hline
 model & $ a_\mu^{\pi^0}\cdot 10^{10}$ & Ref. \\
\hline
\hline\noalign{\smallskip}
 EJLN/BPP & $ 5.9(0.9)$
& \cite{BijnensLBL,Bijnens:2017trn} \\
 Non-local quark model& $ 6.72$
& \cite{Dorokhov:2008pw} \\
 Dyson-Schwinger Eq. Approach & $ 5.75$
& \cite{Goecke:2010if} \\
 LMD+V/KN & $ 5.8-6.3$ 
& \cite{KnechtNyffeler01}\\
 MV: LMD+V+OPE[WZ] & $ 6.3(1.0)$
& \cite{MV03} \\
 Form-factor inspired by AdS/QCD & $ 6.54$
& \cite{Cappiello:2010uy}\\
 Chiral quark model & $ 6.8$
& \cite{Greynat:2012ww}\\
  Magnetic susceptibility constraint & $ 7.2 $
& \cite{Nyffeler:2009tw,JN}\\
\hline
\end{tabular}
\end{table}
\clearpage
Besides the pseudoscalar contributions $\pi^0,\eta,\eta^\prime$ one
similarly can estimate the axial-mesons $a_1,f_1,f_1^\prime$, the
scalars $a_0,f_0,f_0^\prime$, $\pi^\pm,K^\pm$-loops and residual quark-loop
contributions. Tensor mesons~\cite{Pauk:2014rta} 
and a NLO~\cite{Colangelo:2014qya} contribution are also to be included. I then estimate\\[-6mm]
\bea
\amu^{\rm HLbL}&=&[95.5(12.4) +
7.6(2.7)-6.0(1.2)-20(5)+22.3(4)\nn \\
&+&1.1(0.1)+3(2)]\power{-11}=103.4(28.8)\power{-11}\\[-6mm] \nn 
\eea
I have scaled up the quadratically combined error on the l.h.s. by a
factor 2 on the r.h.s. to account for uncertainties which are difficult
to be quantified more precisely. For details I refer to Sect.~5.2.10 of my
book~\cite{Jegerlehner:2017gek}.
\section{Theory vs. Experiment: do we see New Physics?}
Table \ref{tab:thevsexp} compares SM theory with the BNL experimental result.\\[-6mm]
{\small
\begin{table}[h]
\caption{Standard model theory and experiment
comparison in units $10^{-10}$ (see also~\cite{Knecht:2016ygr,Nyffeler:2017ohp,Jegerlehner:2015stw,Jegerlehner:2017lbd}).}
\centering
\begin{tabular}{lr@{.}lr@{.}lc}
\hline
Contribution & \multicolumn{2}{c}{Value} & \multicolumn{2}{l}{Error} & Ref. \\
\hline
QED incl. 4-loops + 5-loops & 11\,658\,471&886 & 0&003 &
\cite{Aoyama12am,Laporta:2017okg,Marquard:2017iib}  \\
Hadronic LO vacuum polarization & 689&46 &  3&25 & \cite{Jegerlehner:2017zsb} \\
Hadronic light--by--light &   10&34 & 2&88 & \cite{Jegerlehner:2017gek}\\
Hadronic HO vacuum polarization & -8&70 & 0&06 & \cite{Jegerlehner:2017zsb} \\
Weak to 2-loops & 15&36 & 0&11 & \cite{Gnendiger:2013pva}  \\
\hline
Theory & \multicolumn{2}{l}{11\,659\,178.3} & \multicolumn{2}{l}{4.3} & --  \\
Experiment & 11\,659\,209&1 & 6&3 & \cite{Bennett:2006fi}  \\
The. - Exp.   4.0 standard deviations &-30&6 & 7&6 & -- \\ \hline
&&&\\[-6mm]
\end{tabular}
\label{tab:thevsexp}
\end{table}}

\noi
What may the 4 $\sigma$ deviation be: new physics?  a
statistical fluctuation?  underestimating uncertainties
(experimental, theoretical)? \textit{Do experiments measure what
theoreticians calculate?} Could it be unaccounted real photon
radiation effects? For possible effects related to lepton flavor violation
see e.g.~\cite{JN,Lindner:2016bgg,Jegerlehner:2017gek} and references therein.
\begin{figure}[t]
\centering
\includegraphics[width=12.5cm]{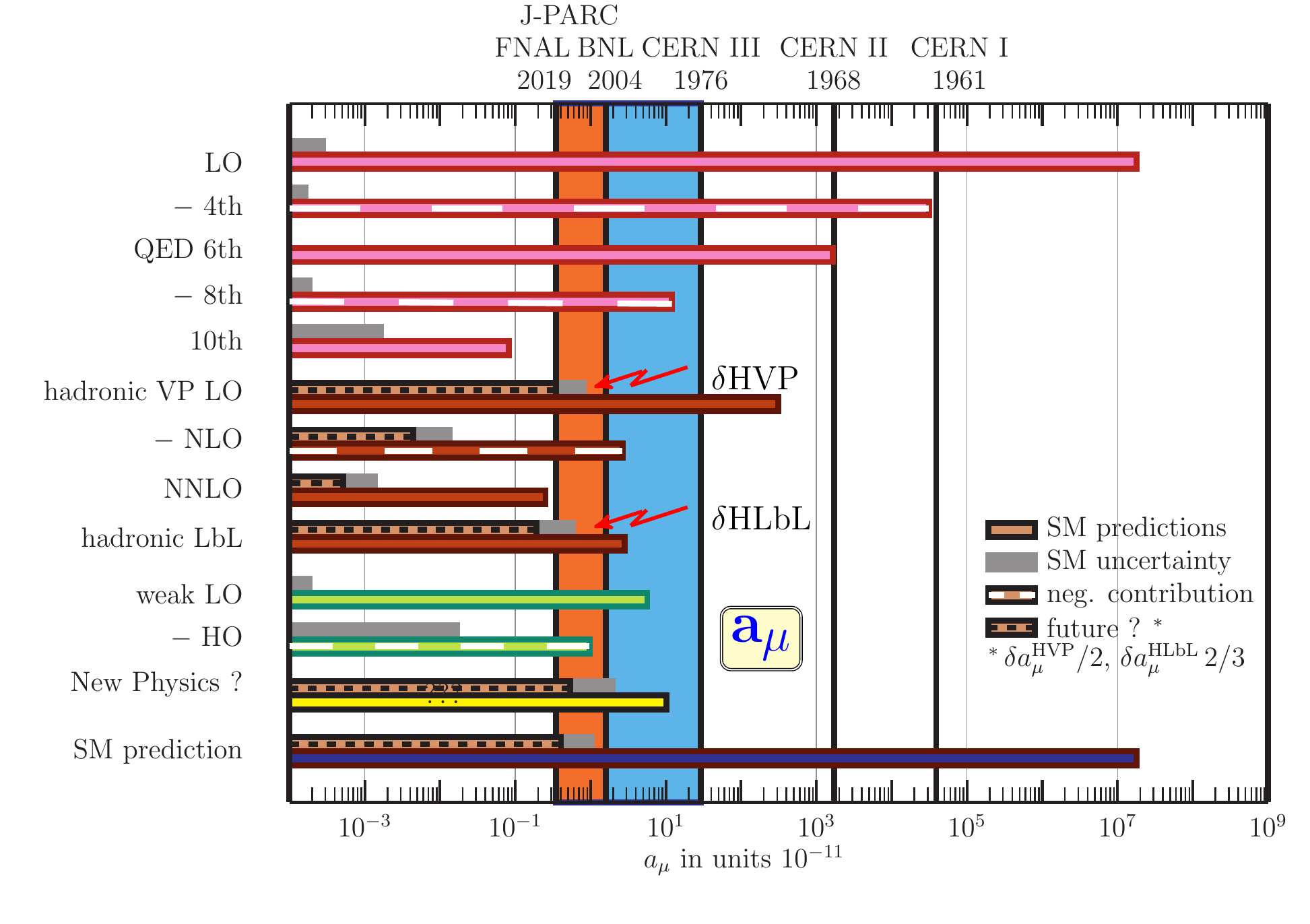}
\caption{Past and future $g-2$ experiments testing various
contributions. New Physics $ \stackrel{?}{=}$ deviation
$ (a_\mu^\mathrm{exp}-a_\mu^\mathrm{the})/a_\mu^\mathrm{exp}$.
Limiting theory precision: hadronic vacuum polarization (HVP) and
hadronic light-by-light (HLbL) (see also~\cite{Mohr:2015ccw}).}
\label{fig:amusensit}
\end{figure}
\begin{figure}[b]
\centering
\includegraphics[width=12.5cm]{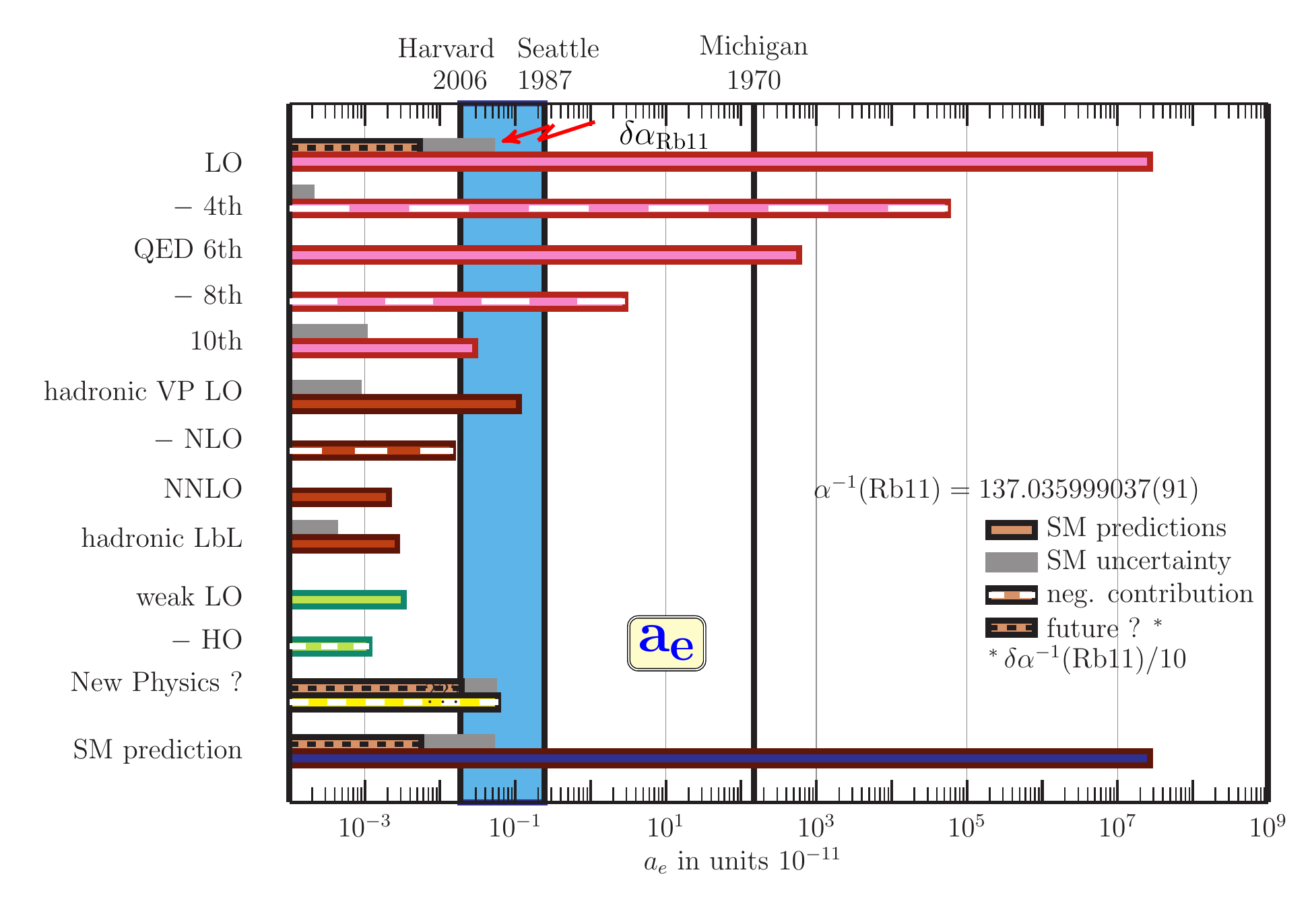}
\caption{Status and sensitivity of the $a_e$ experiments testing various
contributions. The error is dominated by the uncertainty of
$\alpha({\rm Rb11})$ from atomic interferometry. No ``New Physics''
$\stackrel{?}{=}$ deviation
$(a_e^\mathrm{exp}-a_e^\mathrm{the})/a_\mu^\mathrm{exp}$. The blue
band illustrates the improvement by the Harvard experiment.  Note the
very different sensitivities to non-QED contributions in comparison
with $a_\mu$ (for entries see e.g.~\cite{Jegerlehner:2017zsb,Jegerlehner:2017gek,Mohr:2015ccw}).}
\label{fig:aelsensit}
\end{figure}
At the present/future level of precision \mbo{a_\mu} depends on all
physics incorporated in the SM: electromagnetic, weak, and strong
interaction effects and beyond that \textit{all possible new physics}
we are hunting for. Figure~\ref{fig:amusensit} illustrates past and
expected progress in ``the closer we look the more there is to
see''. Here we are and hope to go.  It contrast with the same status
for the electron, Fig.~\ref{fig:aelsensit} shows that $a_e$ still is
and remains a QED test mainly.\\
\noi
{\footnotesize Note added: a new more precise value of $\alpha$ from atomic interferometry with
Cesium$^{133}$ has been obtained at the University of California Berkeley~\cite{alphaCs18}:
$\alpha^{-1}({\rm Cs18})=137.035999046(27)$ giving an $a_e$ prediction
$a_e=0.00115965218157(23)$ such that $a_e^\mathrm{exp}-a_e^\mathrm{the}=(-84\pm36)\power{-14}$
a  $-2.3~\sigma$ deviation. Previously with $\alpha^{-1}({\rm
Rb11})=137.035999037(92)$ we had $a_e=0.00115965218165(77)$
and with $a_e^\mathrm{exp}-a_e^\mathrm{the}=(-92\pm82)\power{-14}$ a
{1.1 $\sigma$} deviation. Although the central value moved closer to
experimental value the deviation has increased owing to the more
precise value of $\alpha$. Note that the
\mbo{a_e} ``discrepancy'' is of opposite sign of the \mbo{a_\mu} one!}


\section{Prospects}
A ``New Physics'' interpretation of the persisting 3 to 4 $\sigma$
gap requires relatively strongly coupled states in the range below about
250 GeV. Search bounds from LEP, Tevatron and specifically from the
LHC already have ruled out a variety of Beyond the Standard Model
(BSM) scenarios, so much hat standard motivations of SUSY/GUT
extensions seem to fall in disgrace. There is no doubt that
performing doable improvements on both the theory and the experimental
side allows one to substantially sharpen (or diminish) the apparent
gap between theory and experiment.

In any case $\amu$ constrains BSM scenarios distinctively and at the
same time challenges a better understanding of the SM prediction.  The
two \underline{complementary} experiments on the way, operating with ultra hot
muons \cite{Grange:2015fou} and with ultra cold muons
\cite{Mibe:2011zz}, repsectively, especially could differ by possible unaccounted real photon
radiation effects. Provided the deviation is real and theory and
needed hadronic cross section data can be improved as expected the muon
\mbo{g-2} experiments could establish \mbo{\Delta \amu^{\rm NP}} at
about \mbo{10} standard deviations. 

A remark concerning HVP issues in the standard data based time-like
approach is in order here:\\
i) How to combine a pretty large number of data-sets to a
truly reliable \mbo{R}-function. What is the true uncertainty? What
part is reliably taken from pQCD? Including or excluding outdated (=older
less precise)  data-sets? Bare versus physical cross
sections, how reliable is VP subtraction?\\
ii) Radiative corrections specifically for the ISR method, sQED issues
etc. The ISR method requires one order in $\alpha$ more precise RC calculation
relative to the SCAN method, at least full 2--loop Bhabha and/or \mbo{\epm
\to \mu^+\mu^-} as well as ISR--FSR interference in the \mbo{\pi^+\pi^-} channel.
What about RC to other more complicated channels (see
e.g.~\cite{CZYZ:2014yra,Jegerlehner:2017kke}). What about
disentangling 30 channels and recombining them in the 1 to 2 GeV
region (quantum interference, missing parts, double counting issues)?\\
iii) What precisely do we need in the DR? The 1PI ``blob'', which
is not a measurable quantity. Need undressing from QED effects, photon
VP subtraction, FSR modeling, \mbo{\rho^0-\gamma} mixing? Do we do
this at sufficient precision?\\
vi) Non-convergence of Dyson series for OZI suppressed narrow
resonances (see e.g.~\cite{Jegerlehner:2015stw}).\\
v) Missing data compatibility among different experiments. Here, global
   fit strategies (see e.g. ~\cite{Benayoun:2011mm,Benayoun:2015gxa})
   can help to learn more about possible problems.

Of course, I think we are doing the best to our knowledge. However, there is no
unambiguous method to combine systematic errors. Uncertainties are
definitely squeezed beyond what can be justified beyond doubts, I think.

Therefore, the very different Euclidean approaches, lattice QCD and the
proposed alternative direct measurements of the hadronic shift $\Delta
\alpha (-Q^2)$~\cite{Abbiendi:2016xup}, in the long term will be
indispensable as complementary cross-checks.
 
For future improvements of the HLbL part one desperately needs more information from
$ \gamma\gamma \to \mathrm{ \ hadrons \ }$ (see e.g.~\cite{Redmer:2018gah})
in order to have better constraints on modeling of the many relevant hadronic amplitudes. 
The dispersive approach to HLbL~\cite{Colangelo:2015ama,Colangelo:2017qdm} is able to allow for real progress since
contributions which were treated so far as separate contributions
will be treated ``rolled into one'' (as entirety). Note that HLbL
depends on 19 independent amplitudes which contribute to $g-2$ while
HVP depends on a single one.
Last but not least: do theoreticians calculate what experiments
measure (form-factor vs cross-section)?
$$a_\mu^{\rm the}= a_\mu^{\rm SM \ virtual}{ \:[=F_2(0)]}+ \underbrace{\Delta a_\mu^{{\rm SM
\ real \ soft \ } \gamma}}_{???} \:\underbrace{ \mathrm{[dep. \ on
\ exp. \ setup]}}_{\rm Fermilab \ vs \ J-PARC}
+ \Delta a_\mu ^{\rm NP}\epo$$
A lot remains to be done while a new $a^{\rm exp}_\mu$ is in sight.

\noindent
{\bf Acknowledgments} \\ Many thanks to the organizers for the kind
invitation to the 2018 Cracow EPIPHANY Conference and for giving me the opportunity to present
this talk. I gratefully acknowledge the support by DESY.

\end{document}